\documentclass[reprint,superscriptaddress,showpacs,nofootinbib,aps,prb]{revtex4-1}
\usepackage{amssymb}
\usepackage{amsmath}
\usepackage{graphicx}
\usepackage{dcolumn}
\usepackage{bm}
\usepackage[utf8]{inputenc}

\begin{document}

\title{Electric-field control of spin transitions in molecular compounds}
\author{O.\ I.\ Utesov}
\affiliation{National Research Center ``Kurchatov Institute'' B.P.\ Konstantinov Petersburg Nuclear Physics Institute, Gatchina 188300, Russia}
\affiliation{St.\ Petersburg State University, 7/9 Universitetskaya nab., St.\ Petersburg 199034, Russia}
\affiliation{St. Petersburg Academic University - Nanotechnology Research and Education Centre of the Russian Academy of Sciences, 194021 St.\ Petersburg, Russia}
\author{S. Burdin}
\affiliation{University of Bordeaux, LOMA UMR-CNRS 5798, F-33405 Talence Cedex, France}
\author{P. Rosa}
\affiliation{CNRS, Univ. Bordeaux, Bordeaux INP, ICMCB, UMR 5026, F-33600 Pessac, France}
\author{M. Gonidec}
\affiliation{CNRS, Univ. Bordeaux, Bordeaux INP, ICMCB, UMR 5026, F-33600 Pessac, France}
\author{L. Poggini}
\affiliation{CNRS, Univ. Bordeaux, Bordeaux INP, ICMCB, UMR 5026, F-33600 Pessac, France}

\author{S. V. Andreev}
\email[Electronic adress: ]{Serguey.Andreev@gmail.com}
\affiliation{ITMO University, Saint-Petersburg, Russia}

\date{\today}

\begin{abstract}
We present a theoretical model of spin transitions in stacks of molecular layers. Our model captures the already established physics of these systems (thermal hysteretic transitions and crossovers) and suggests a way towards in situ control of this physics by means of an external electric field. Our results pave the way toward both temperature and voltage controllable organic memory.
\end{abstract}

\pacs{Valid PACS appear here}
\maketitle

\preprint{APS/123-QED}

\section{Introduction}
\label{intro}

Spin crossover molecular compounds (SCO) have been intensively discussed during the past decades in connexion with data recording and sensing \cite{Kahn, Felix}. These systems switch between a low (LS) and a high-spin (HS) state as the temperature is increased. In some cases the conversion has a hysteretic behaviour characterized by heating and cooling characteristic temperatures, $T_c^{\uparrow}$ and $T_c^{\downarrow}$, respectively. These temperatures can be easily made on the order of the room temperature. By applying temporary heating (e.g., a laser pulse) one can switch an initially LS state to a HS state. The latter will be preserved until cooling the device below the operational temperature.

At the microscopic level, the conversion is due to switching between two electronic states of molecules characterized by different occupation of $e_g$ and $t_{2g}$ subsets of $3d$ metal orbitals. The LS state arises from the closed-shell ($t_{2g}^6$) and the HS state from the open-shell ($t_{2g}^{4}e_g^2$) configurations  \cite{Kahn}. These differ by magnetic, optical and structural properties and can be altered by pressure, temperature and light irradiation \cite{Gutlich1, Halcrow, Bousseksou1, Gutlich2} which makes SCO promising for new functional materials \cite{Feringa}. SCO complexes consist of transition metal ions surrounded by organic ligands. One may play with the SCO thermodynamics by carefully designing the ligands. At the spin crossover, the enthalpy remains essentially constant with temperature and the SCO phenomenon is driven by entropy. For the HS state the electronic contribution to entropy is higher than that of the LS state. As the SCO needs to be accompanied by a structural change of the complex resulting in weaker bonds in the HS state, the vibrational contribution to entropy for the latter is also higher than for the LS state. This situation leads to a thermal conversion from the LS state to HS state upon increasing temperature \cite{Lefter}.

The most transparent theoretical description of the SCO physics, as demonstrated in numerous works \cite{Koudriavtsev, Wajnflasz, Bari, Zelentsov, Bousseksou2, Kamel}, can be done in terms of the Ising-like model:
\begin{equation}
  \label{Ising}
  \hat H= \Delta\sum\limits_{i=1}^N \hat S_z^i-J\sum_{i,j} \hat S_z^i \hat S_z^j,
\end{equation}
with the relevant parameters being the energy splittings between the LS and HS states $2\Delta$ , and the ferromagnetic-like coupling constant $J>0$ describing the interaction between the nearest neighbours (``cooperativity effect''). Due to degeneracy of the open-shell $t_{2g}^{4}e_g^2$ electronic configuration the HS state has larger statistical weight and thus stabilizes at sufficiently large temperatures.

The situation is less understood for thin films of SCO molecules. The studies of SCO films with thicknesses ranging from 5 to 1000 nm conclude that the thermally driven spin transition in such systems is similar to that of the bulk \cite{Lefter, Senthil, Naggert, Shi, Palamarciuc, Gruber, Mahfoud, Baadji, Lorenzo1, Lorenzo2}. However, when the thickness is decreased down to sub-monolayer or a few monolayers in coverage, the SCO behavior seems to be modified by the interaction with the substrate \cite{Gopakumar1, Gopakumar2, Pronschinske, Bernien, Warner, Lorenzo3, Barreteau}. In particular, some of us have recently demonstated that a thick film of [Fe(H$_2$B(pz)$_2$)$_2$(bipy)] deposited by thermal sublimation on an organic ferroelectric substrate maintains the SCO behaviour \cite{Zhang1}, whereas for thinner films (under 15 nm) the SCO behavior was controlled by the substrate polarization \cite{Zhang1, Zhang2, Lorenzo4}.

On the general grounds, one may expect the substrate to modify the splitting $\Delta$ between the spin states of the molecules. First, the splittings of molecules which constitute the boundary layer are clearly affected by microscopic Van-der-Waals interaction with the surface of the substrate. This effect seems to be responsible for the recent experimental observations \cite{Zhang1, Zhang2, Barreteau}. Second, macroscopic electric field $\bm E$ produced by the ferroelectric substrate should modify the splittings in all layers according to the formula
\begin{equation}
  \label{Coupling}
  \Delta(\bm E)=\Delta_0+\frac{1}{2}\sum_{\alpha,\beta=x,y,z}\upsilon_{\alpha\beta} E_\alpha E_\beta,
\end{equation}
where we assume the field being uniform over the sample, $\Delta_0$ is the bare splitting at $E=0$ and $\upsilon_{\alpha\beta}$ is some phenomenological \textit{symmetric} tensor to be defined from the experiment. Existence of the coupling of the type \eqref{Coupling} may be argued as follows. Due to the electrostriction the electric field would modify the pressure $P$ acting on the system (or, alternatively, the system volume $V$). The change in the pressure $P$ can be written as
\begin{equation}
\label{electrostriction}
\delta P=\frac{1}{2}\sum_{\alpha,\beta}\left[\frac{\partial (\alpha_{\alpha\beta}V)}{\partial V}\right]_T E_\alpha E_\beta,
\end{equation}
where $\alpha_{\alpha\beta}$ is the polarizability of the sample.\cite{Landau} This change of the pressure, on the other hand, would modify the spin splittings due to an inverse "magnetostriction" effect,
\begin{equation}
\label{magnetostriction}
\Delta=\Delta_0+\left(\frac{\partial\Delta}{\partial P}\right)_{\delta P=0}\delta P.
\end{equation}
What we call "magnetostriction" in the context of the model \eqref{Ising} is a phenomenological way to account for the fact that the average metal-ligand bond length is longer in the HS state than in the LS state \cite{Volume}. For instance, the characteristic temperature of the crossover $T_{1/2}$ was shown to grow linearly with the pressure.\cite{Pressure1,Pressure2} As we shall see, this experimental fact justifies the assumption \eqref{magnetostriction} \textit{a posteriori} and, therefore, supports existence of the relation \eqref{Coupling}. 

Implementation of the coupling \eqref{Coupling} would build a bridge between the field of spin transition polymers and ferroelectricity, and pave a way toward both temperature and voltage controllable organic memory. A crucial first step on this way is a theoretical analysis of implication of the hypothesis of a tunable $\Delta$ on the physics of spin transitions as described by the Hamiltonian \eqref{Ising}. This analysis is presented in our paper. We start by formulating the theoretical model we use in the present study. In Sec.~\ref{theor}, under certain assumptions, we obtain the effective Hamiltonian of the system in the form~\eqref{Ising}. For the bulk problem its mean-field solution is given in Sec.~\ref{one-layer}. The already established physics of thermal spin transitions and crossovers is presented in Subsec.~\ref{temp}. The main result of this subsection is the existence of the critical value of the ratio $\Delta/J$ above which the first-order thermal transition (hysteresis) turns to a smooth crossover. We obtain a simple analytical expression for this ratio and confirm it by numerics. In Subsec.~\ref{delta} we show that isothermal variation of $\Delta$ can also lead to a hysteresis. Arguments, analogous to those presented in Subsec. III A, when applied to the spin transition induced by variation of $\Delta$, yield the maximum and minimum temperatures at which the hysteresis under an electric field would be possible.

In Sec.~\ref{multi-layer} we turn to the discussion of layered systems. We show that, for sufficiently weak coupling between the layers, it should be possible to observe a staircase in the total magnetization as a function of $\Delta$. We discuss such \textit{multistability} in terms of a phase diagram of $\Delta$ versus the ratio of inter- to intra-layer couplings for two layers. For sufficiently large inter-layer coupling the transition occurs in both layers simultaneously. This switching can be performed either as in bulk (by varying either temperature or $\Delta$) or by variation of the energy splitting $\Delta$ only in the boundary layer (due to, e.g., interaction with the surface of the substrate). This result of our theoretical model holds qualitatively for few layers (thin films). In contrast, in films with large number of layers, variation of $\Delta$ in the boundary layer has no impact on the total magnetization. The situation here is similar to the bulk, in agreement with the experiment \cite{Zhang1, Zhang2}.

\begin{figure}[b]
  \includegraphics[width=1\columnwidth]{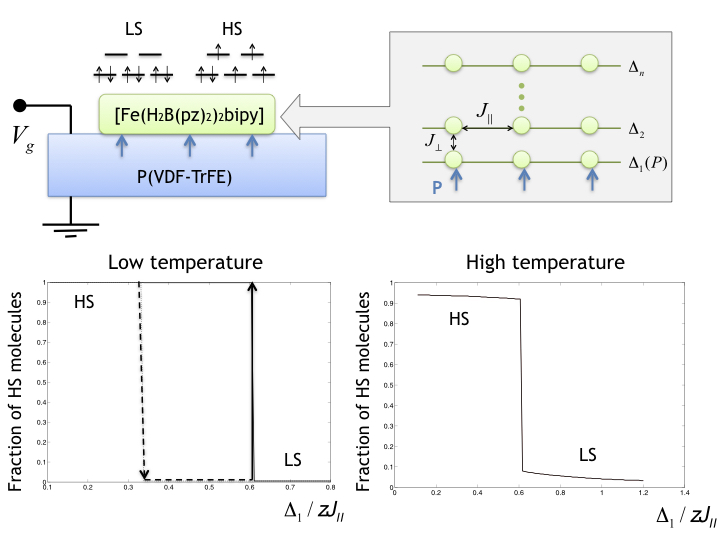}
  \caption{Switching of the spin state of a stack of $n$ molecular thin films [Fe(H$_2$B(pz)$_2$)$_2$bipy] (green rectangle) by the polarization of a ferroelectric substrate (blue rectangle). Each molecule can be in two possible states, "low spin" (LS) and "high spin" (HS) characterized by zero and non-zero spin projection on the growth axis, respectively. To describe the behavior of the system we use the model shown in the gray rectangle on the right (see the text). In the bottom panel we show the fraction of HS molecules as a function of the HS/LS splitting at the boundary layer (defined by the polarization $P$ of the substrate) for two different temperatures $T$. At low $T$ one has a first order transition (hysteresis) which turns to a crossover at high $T$. We used the model \eqref{dfen1} with $n=2$, $\sigma=0.5$ and the following set of parameters: (i) $t=0.5$, $\delta_2=0.2$ and (ii) $t=1$, $\delta_2=1$.}
  \label{Fig0}
\end{figure}

\section{Theoretical model and methods}
\label{theor}

Our starting point is the following phenomenological Hamiltonian
\begin{equation}
  \label{Hamiltonian1}
  \hat H=\sum_i \hat H_i+\sum_{i,j}\hat V_{ij}
\end{equation}
where
\begin{equation}
  \hat H_i= E_0^i |0\rangle_i \langle0|_i+E_1^i \sum_{\alpha=1..g} |1_\alpha\rangle_i \langle 1_\alpha|_i
\end{equation}
and
\begin{equation}
  \hat V_{ij}=\sum_{\sigma_1,\sigma_2,\sigma_3,\sigma_4}J_{\sigma_1,\sigma_2,\sigma_3,\sigma_4,i,j}|\sigma_1\rangle_i\otimes |\sigma_2\rangle_j\langle\sigma_3|_j\otimes\langle\sigma_4|_i.
\end{equation}
The Hamiltonian \eqref{Hamiltonian1} is a general model of a lattice of two-level systems with interaction. We attribute $|0\rangle_i$ to the low-spin (LS) and $|1_\alpha\rangle_i$ to the one of the $g$-fold degenerate high-spin (HS) states of the $i$-th molecule, respectively. By introducing the pseudo-spin operators
\begin{equation}
\begin{split}
  \hat S_z&=\sum_\alpha |1_\alpha\rangle\langle1_\alpha|-|0\rangle\langle0|,\\
  \hat S_+&=\hat S_x+i\hat S_y=\sum_\alpha |1_\alpha\rangle\langle0|,\\
  \hat S_-&=\hat S_x-i\hat S_y=\sum_\alpha |0\rangle\langle1_\alpha|,
\end{split}
\end{equation}
we rewrite the Hamiltonian \eqref{Hamiltonian1} in the useful form
\begin{equation}
\label{Hamiltonian2}
\hat H=\sum_i \Delta_i \hat S_z^i-\sum_{i,j}\sum_{\mu,\nu=(+,-,z)}\hat S_\mu^i M_{\mu\nu}\hat S_\nu^j
\end{equation}
where we have introduced $\Delta_i=(E_1^i-E_0^i)/2$ for the molecular energy splitting. The crucial assumption now is to take the second term in \eqref{Hamiltonian2} in the block form $\hat V_{ij}=- (M_{zz} \hat S_z^i \hat S_z^j+M_{+-} \hat S_+^i \hat S_-^j+M_{-+} \hat S_-^i \hat S_+^j)$ and assume the interaction between the nearest neighbors only. The problem is thus projected onto an effective Heisenberg XXZ model. The first term in the above equation describes the static interaction between the spins, whereas inclusion of the second term would allow to study the dynamical response to perturbations. In this work we shall examine the case $M_{zz}=J>0$ (ferromagnetic-like coupling) and use the static mean-field approximation $<\hat S_z^i>=m$ and $<\hat S_\pm^i>=0$. Thus, the model is reduced to an effective Ising model~\eqref{Ising} with ``magnetic field'' $\Delta$.

To take into account the structure of the system (the layers are arranged on the top of each other along the growth direction) we shall further introduce $J_\parallel$ for the interlayer coupling and $J_\perp$ to describe interaction between the layers. The quantities $z_\perp$ and $z_\parallel$ will be the corresponding coordination numbers (see Fig. \ref{Fig0}).

After the model Hamiltonian has been constructed, all the relevant thermodynamic quantities can be obtained starting from the partition function
\begin{equation}
\label{Z}
Z=Tr(e^{-\beta \hat H}),
\end{equation}
where $\beta=1/k_B T$ is the inverse temperature of the system. The free energy reads
\begin{equation}
F=-k_B T \ln (Z)
\end{equation}
and, considered as a function of the average magnetization $m$, can be used to describe transitions between different states, as we show below.

\section{Single layer}
\label{one-layer}

\subsection{Thermal hysteresis}
\label{temp}

It is instructive to discuss first the simplest case of a homogeneous single-layer system (which is, of course, physically equivalent to the bulk). The Hamiltonian reads
\begin{equation}
\hat H_1=\Delta\sum\limits_{i=1}^N \hat S_z^i-\sum_{i,j}J \hat S_z^i \hat S_z^j,
\end{equation}
where the summation in the second term is over the nearest neighbors and $N$ is the total number of molecules in the layer. By introducing the coordination number $z$ (the number of the nearest neighbors) and using the mean-field approximation we rewrite the above Hamiltonian in the form
\begin{equation}
\hat H_1=\sum\limits_{i=1}^N (\Delta-J z m)\hat S_z^i+\frac{1}{2}\sum\limits_{i=1}^N J z m^2,
\end{equation}
where $Jzm$ is an effective Weiss field. Using Eq. \eqref{Z}, we calculate the free energy
\begin{equation}
\frac{F}{N}=\frac{Jzm^2}{2}-k_B T\ln \left[ g e^{-(\Delta - J z m)/k_B T}+e^{(\Delta - J z m)/k_B T} \right].
\end{equation}
Here $g$ is the degeneracy of the HS state, discussed above.

It is convenient to introduce the dimensionless parameters $t=k_B T/J z $ and $\delta=\Delta/J z$. The dimensionless free energy per molecule then reads
\begin{equation} \label{fen1}
f(m)=\frac{m^2}{2}-t\ln\left[ge^{-(\delta-m)/t}+e^{(\delta-m)/t}\right].
\end{equation}
Considered as a function of $m$ the free energy can have either two minima separated by a barrier or one minimum. The latter situation is always realized at sufficiently large temperatures $t$. At moderate temperatures there can be two physically distinct scenarios depending on the value of $\delta$: first order transition with hysteresis and a cross-over.

In order to describe these two scenarios analytically we first consider the low-temperature limit $t \ll 1$ and $\delta \sim t$, where simple analytical expressions can be derived. Their region of validity will be discussed below. Near the $m=1$ point one can neglect the second exponent in the logarithm in Eq.~\eqref{fen1} and obtain the following form for the free energy :
\begin{equation} \label{fen2}
f(m)=\frac{m^2}{2}-m - t\ln{g} + \delta,
\end{equation}
which evidently yields a local minimum $m=1$. Analogously, near $m=-1$,
\begin{equation} \label{fen3}
f(m)=\frac{m^2}{2}+m  - \delta,
\end{equation}
and there is a local minimum at $m=-1$. There is also a maximum at $m\approx 0$. So, the properties of the system are defined by three free energies
\begin{equation} \label{fen4}
\begin{split}
   f(+1)&\approx-1/2 - t \ln g + \delta,\\
   f(-1)&\approx-1/2 - \delta,\\
   f(0)&=-t \ln{\left(g e^{-\delta/t} + e^{\delta/t} \right)}.
\end{split}
\end{equation}
Obviously, at very low temperature the system will be in the LS state. Then, from the condition $f(+1)=f(-1)$ we can determine the temperature at which the ground state become doubly degenerate (we shall refer to this temperature as $t_{1/2}$). Simple calculation yields
\begin{equation}\label{t12}
  t_{1/2}=\frac{2\delta}{\ln{g}}.
\end{equation}
Then, two cases should be distinguished. The barrier height $h(t)$ at the temperature $t_{1/2}$ can either exceed or be lower than this temperature. In the first case the thermal fluctuations at $t=t_{1/2}$ are insufficient to induce a transition from LS to HS state. One has to attain some larger temperature $t_{\uparrow}$ at which $h(t_{\uparrow})=t_{\uparrow}$ in order to observe the transition. On the other hand, when decreasing $t$, an inverse transition from HS to LS state apparently cannot take place at  $t_{\uparrow}$, so that one should define some $t_{\downarrow}$ such that $t_{\downarrow}<t_{\uparrow}$, and a natural way to do it is to let $t_{\downarrow}\equiv t_{1/2}$. Using equations above we can find the critical value $\delta_{cr}$ which separates two different regimes:
\begin{equation}\label{dcr1}
  h(t_{1/2})= \left. f(0)-f(-1) \right|_{t=t_{1/2}} = t_{1/2},
\end{equation}
which solution gives
\begin{equation}\label{dcr2}
  \delta_{cr} = \frac{\ln{g}}{4(1+\ln{2})}.
\end{equation}
We consider $g=5$, so $\delta_{cr} \approx 0.24$. For $\delta<0.24$ the barrier height $h(t_{1/2})>t_{1/2}$ and one has a first order thermal spin transition characterized by a hysteresis loop. Indeed, after simple calculations one can derive equation for $t_\uparrow$:
\begin{equation}\label{tup}
  \delta = -\frac{t_\uparrow}{2} \ln{\left[ \frac{e^{\frac{1}{2 t_\uparrow}-1}-1}{g}\right]},
\end{equation}
which yields $t_\uparrow > t_{1/2}$ at $\delta < \delta_{cr}$. We check the formulas~\eqref{t12},~\eqref{dcr2} and~\eqref{tup} numerically and find that they hold with excellent accuracy in the whole range of the relevant values of $\delta \lesssim \delta_{cr}$.

\begin{figure}[t]
\includegraphics[width=1\columnwidth]{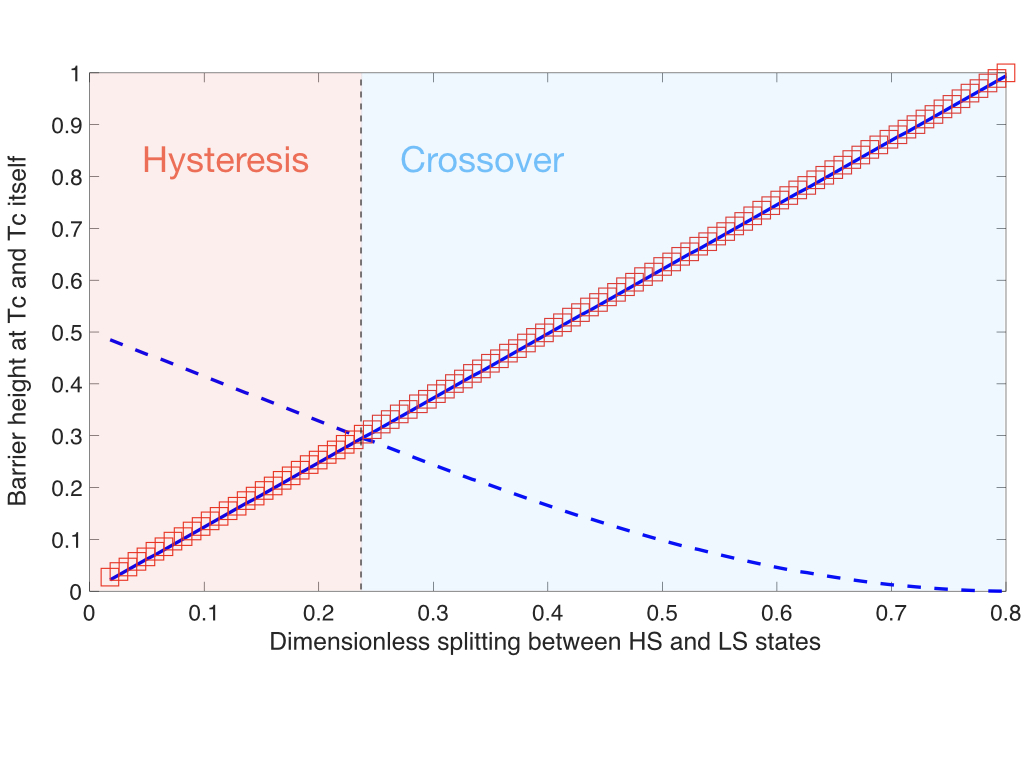}
\caption{Dependence of the barrier height $h(t_{1/2})$ (dashed line) and the characteristic temperature $t_{1/2}$ (open squares and solid line) on the parameter $\delta$. The critical value $\delta=0.24$ is defined as a point at which  $h(t_{1/2})=t_{1/2}$. For $\delta<0.24$ one has a regime of hysteresis (left area, see the text). For $\delta>0.24$ there is a smooth crossover between LS and HS states and vice versa along the same curve. For the $t_{1/2}$ dependence the open squres are used for the numerical result, whereas the solid line is the analytic expression \eqref{t12}. Note an excellent agreement between the two.}
\label{Fig1}
\end{figure}

For $\delta>\delta_{cr}$ the situation is very different. Now one has $h(t_{1/2})<t_{1/2}$, which means that the first order transition is replaced by a smooth crossover from LS to HS and vice versa along the same curve. The two distinct regimes are shown by red ($\delta<0.24$) and blue ($\delta>0.24$) colors in Fig. \eqref{Fig1}.

At $\delta\approx 0.75$ the barrier at $t=t_{1/2} \approx 0.93$ disappears, which means that the potential relief almost flattens and the two minima of $f(m)$ merge. Below, we will use the value $t_{1/2}=0.93$ to quantify the situation where in the equation
\begin{equation}\label{min1}
  m = \tanh{\frac{m}{t_{1/2}}},
\end{equation}
(which is the exact equation for extrema of the free energy~\eqref{fen1} at $t_{1/2}$~\eqref{t12}) the distinct minima at $m \approx \pm 1$ disappear.

To close this subsection we notice that the obtained linear growth of $t_{1/2}$ with $\delta$ corroborates \textit{a posteriori} the assumption \eqref{magnetostriction} of the linear dependence of $\Delta$ on the pressure. Indeed, linear growth of $T_{1/2}$ with the applied pressure has been previously reported in the experimental studies.\cite{Pressure1,Pressure2} 

\subsection{Isothermal switching}
\label{delta}

In the previous subsection we have shown that the transition temperature is determined by the parameter $\delta$ --- energy splitting between two states (see Eqs.~\eqref{t12} and~\eqref{tup}). A new possibility of layer state switching arises from Eq.~\eqref{Coupling}. Naturally, we can ``inverse'' the above analysis. To do so, we can fix temperature and vary $\delta$ around
\begin{equation}\label{dvar}
  \delta(t) \equiv \frac{ t \ln{g}}{2}.
\end{equation}
At such $\delta(t)$, LS and HS states are energetically equivalent and variation of $\delta$ evidently leads to isothermal switching between LS and HS states.

Let us, for instance, start from the LS state at the temperature $t=t_{1/2}(\delta)$. Then, if we change $\delta$ to $\delta_\uparrow<\delta(t)$ such as the barrier height becomes small enough ($\leq t $) switching to HS state will occur. The latter condition is satisfied if $t_{1/2}(\delta)>t_\uparrow(\delta_\uparrow)$. If one then returns back to $\delta(t)$ the system will remain in the HS state. The required difference can be expressed using Eqs.~\eqref{t12} and~\eqref{tup} as
\begin{equation}\label{dvar2}
  \delta_\uparrow-\delta = - \frac{\delta}{\ln{g}} \ln{\left[ e^{\ln{g}/4 \delta -1 } -1\right]}.
\end{equation}
It is reasonable to consider the situation where we can not change the molecular energy levels significantly, and the LS state should have lower energy than the HS state. This is equivalent to condition $\delta_\uparrow > 0$, which yields
\begin{equation}\label{dvar3}
  \delta(t)>\frac{\ln{g}}{4[\ln{(g+1)}+1]} \approx 0.14.
\end{equation}
should hold. Using Eq.~\eqref{dvar} it can be rewritten as the restriction on the temperature:
\begin{equation}\label{dvar4}
  t >\frac{1}{2[\ln{(g+1)}+1]} \approx 0.18.
\end{equation}
At lower temperatures $\delta$-variation is insufficient to induce transition to the HS state because the barrier is too high. Moreover, the higher the temperature the lower $\delta$ variation is required to perform switching.

To switch the state back to LS one should increase $\delta$ to some $\delta_\downarrow>\delta(t)$ which makes LS state energetically preferable and barrier height smaller than the system temperature. After some calculations we obtain
\begin{equation}\label{dvar5}
  \delta_\downarrow-\delta = \frac{\delta}{\ln{g}} \ln{\left[ e^{\ln{g}/4 \delta -1 } -1\right]}.
\end{equation}
We should also require $ \delta_\downarrow-\delta(t)>0$ in order to have the hysteresis. This condition is equal to $\delta(t)<\delta_{cr}$ and (see Eqs.~\eqref{t12} and~\eqref{dcr2})
\begin{equation}\label{dvar6}
  t< \frac{2 \delta_{cr}}{\ln{g}} \approx 0.29.
\end{equation}

Thus, in the certain interval of parameters, we obtain the hysteresis which is controlled by the energy splitting between LS and HS state, not by the temperature.

\begin{figure}
  \centering
  \includegraphics[width=1\columnwidth]{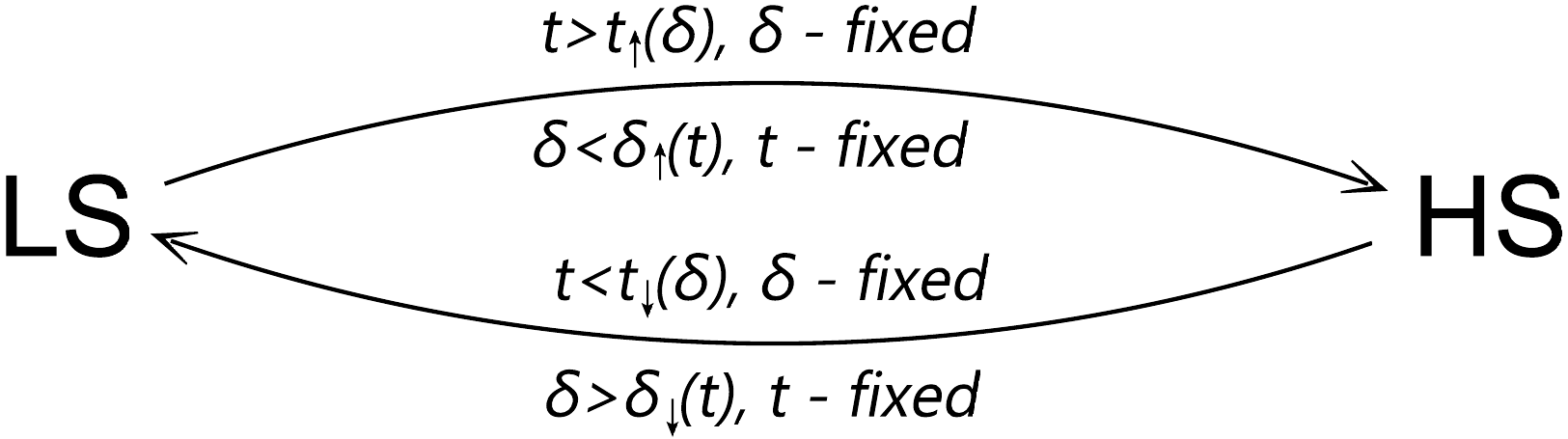}\\
  \caption{Ways to switch the spin state of the system. In a certain range of parameters (see the main text) the transitions are characterized by hysteresis and can be accomplished by varying either the temperature or the energy splitting between the LS and HS states.}\label{Scheme}
\end{figure}

We summarize our analysis of the single-layer problem in Fig.~\ref{Scheme}.

\section{Multilayer problem. Multistability}
\label{multi-layer}
After having established the fundamentals of the thermal hysteresis/crossover in bulk, we now turn to investigation of a layered structure deposited on a ferroelectric substrate. As we have conjectured in the Introduction, the substrate primarily affects the value of the molecular energy splitting $\Delta_1$ at the boundary layer. In the frame of our model such coupling can be described by Eq. \eqref{magnetostriction}, where the pressure $P$ is due to boundary effects (e.g., the epitaxial strain\cite{Barreteau}). Clearly, such boundary strains are also affected by the substrate polarization. However, in contrast to the bulk problem, here one cannot use the macroscopic relation \eqref{electrostriction}. We therefore merely assume a possibility of tuning $\Delta_1$, leaving the detailed investigation of the underlying mechanisms for future studies. Increasing the ratio
\begin{equation}
\sigma=\frac{J_\perp}{z_\parallel J_\parallel},
\end{equation}
drives the system to a cooperative regime, where the change in $\Delta_1$ results in switching of the spin state of the whole sample (simultaneous transition from LS to HS in both layers). To develop a feel of the effect it is instructive to consider first the case of two layers.

\subsection{Two layers}
\label{bilayer}
\begin{figure}[t]
\includegraphics[width=1\columnwidth]{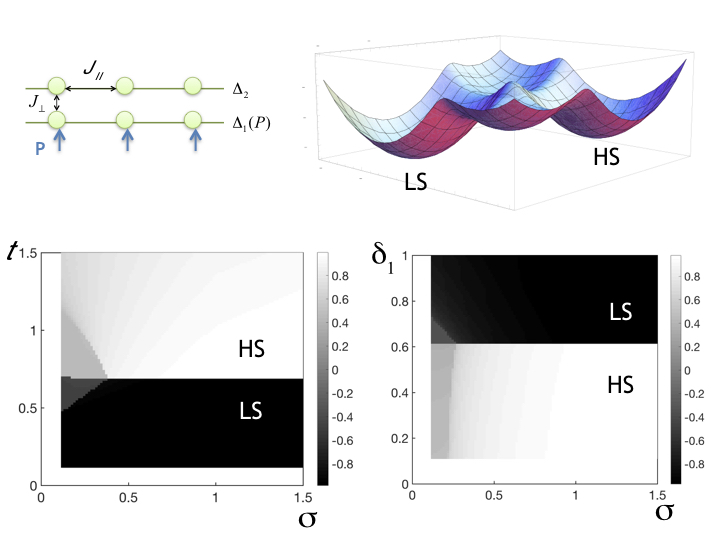}
\caption{(Top) Schematic illustration of a two-layer setting and the corresponding free energy landscape $f(m_1,m_2)$. The labels LS and HS denote the areas in the $(m_1,m_2)$ plane where the entire system is in the low- /high-spin states, respectively. (Bottom) $(\sigma,t)$ (left, $\delta_1=0.1$) and $(\sigma,\delta_1)$ (right, $t=1$) phase diagrams of a double-layer system. We take $\delta_2=1$. The color is the average magnetization $m=(m_1+m_2)/2$: black corresponds to $m=-1$ and white to $m=+1$. For sufficiently large $\sigma$ one can switch the spin state of the whole system by varying $\delta_1$ near some critical value $\delta_{1c}$.}
\label{Fig2}
\end{figure}

The dimensionless free energy of the double-layer system reads
\begin{equation}\label{dfen1}
  \begin{split}
  &f(m_1,m_2)=\frac{1}{2}\left(m_1^2+m_2^2+2\sigma m_1 m_2\right)\\
  &-t\left[\ln\left(ge^{-e_1/t}+e^{e_1/t}\right)+\ln\left(ge^{-e_2/t}+e^{e_2/t}\right)\right],
  \end{split}
\end{equation}
where
\begin{equation}
e_{1,2}=\delta_{1,2}-m_{1,2}-\sigma m_{2,1}
\end{equation}
and $g=5$ accounts for the 5-fold degeneracy of the HS state. We have introduced the following notations: $f=F/NJ_\parallel z_\parallel$, $N$ being the number of molecules in one layer, $t=k_BT/J_\parallel z_\parallel$, $\delta_\alpha=\Delta_\alpha/J_\parallel z_\parallel$.

We start from the similar to the previous section analysis of the free energy at low temperatures. We shall refer to states with different $m_1$ and $m_2$ as $(m_1,m_2)$.

Near $(+1, +1)$ point we have the free energy in the form
\begin{equation}
  \begin{split}
  f &\approx \frac{m^2_1+m^2_2 + 2 \sigma m_1 m_2}{2} -(1+\sigma)(m_1+m_2) \\ &-2t \ln{g} + (\delta_1+\delta_2).
  \end{split}
\end{equation}
This function evidently has minimum at $(+1,+1)$. The same analysis can be also performed for $(-1, -1)$, $(-1, +1)$ and $(+1, -1)$ points. It shows that $(-1,-1)$ is a local minimum. However, two other points $(-1,+1)$ and $(+1,-1)$ can be minima only if $\sigma<\sigma_{c}<1$, where $\sigma_{c}$ is dependent on $t,\delta_1,\delta_2$ and will be determined below. Corresponding free energies read
\begin{eqnarray}\label{dfen2}
  f(-1,-1) &\approx& -1 - \sigma  - (\delta_1+\delta_2),  \\ \label{dfen3}
  f(+1,+1) &\approx&  -1 - \sigma - 2t \ln{g} + (\delta_1+\delta_2).
\end{eqnarray}
At $\sigma<\sigma_{c}$ we have additional minima with free energy
\begin{eqnarray}\label{dfen4}
  f(+1,-1) &\approx& \sigma -1 - t \ln{g}  + \delta_1 - \delta_2,  \\ \label{dfen5}
  f(-1,+1) &\approx&  \sigma -1 - t \ln{g}  + \delta_2 - \delta_1.
\end{eqnarray}
These additional minima make the transition from the LS in both layers $(-1,-1)$ to HS $(+1,+1)$ indirect, for example sequence $(-1,-1)\rightarrow(+1,-1)\rightarrow(+1,+1)$ can arise. We shall refer to this case as \textit{multistability}.

It is seen from the equations above, that at low temperatures $(-1,-1)$ state is the global minimum. At
\begin{equation}
  \label{dt12}
  t_c^{(2)}=\frac{\delta_1+\delta_2}{\ln g}
\end{equation}
we have $f(+1,+1)=f(-1,-1)$, thus it is the temperature when the ground state is doubly degenerate.

We also notice a possibility for $(+1,-1)$ or $(-1,+1)$ to be a global minimum in some temperature interval of temperatures near $t_c^{(2)}$. Indeed, one can see from Eqs.~\eqref{dfen2} and~Eqs.~\eqref{dfen4} that if $2 \sigma < \delta_1 - \delta_2$ then $f(-1,+1) < f(-1,-1)$ at $t_c^{(2)}$. 

At high enough interlayer interaction ($\sigma>\sigma_c$) the transition from LS to HS occurs simultaneously in both layers. In this case the minima at $(+1,-1)$ and $(-1,+1)$ disappear in the critical temperature $t_c^{(2)}$ vicinity. In order to quantify $\sigma_c$ we rewrite~\eqref{dfen1} using new variables, $m_I = m_1 + \sigma m_2$ and $m_{II} = m_2 + \sigma m_1$. We obtain the free energy in the form
\begin{equation}
  \begin{split}
  f(m_1,m_2)&=\frac{m_I^2+m_{II}^2- 2\sigma m_I m_{II}}{2(1-\sigma^2)}\\
  &-t\Bigl[\ln(ge^{(m_I-\delta_1)/t}+e^{(\delta_1-m_I)/t}) \\
  &+\ln(ge^{(m_{II}-\delta_2)/t}+e^{(\delta_2-m_{II})/t})\Bigr].
  \end{split}
\end{equation}
In a special case of $\delta_1=\delta_2$ at $t_c^{(2)}$ the system of equations which defines free energy minima reads (cf. Eq.~\eqref{min1})
\begin{eqnarray}
\left\{
  \begin{array}{c}
    \frac{m_I-\sigma m_{II}}{1 - \sigma^2} = \tanh{(m_I/t_c^{(2)})}, \\
    \frac{m_{II}-\sigma m_{I}}{1 - \sigma^2} = \tanh{(m_{II}/t_c^{(2)})}. \\
  \end{array}
\right.
\end{eqnarray}
At small enough $t_c^{(2)}$ all the right hand sides of these equations can be substituted by $\pm 1$ and the system gives previously discussed solutions, but written in new variables: $(-1-\sigma,-1-\sigma)$, $(1+\sigma,1+\sigma)$, $(1-\sigma,\sigma-1)$ and $(\sigma-1,1-\sigma)$. From Eq.~\eqref{min1} we saw that if the $\tanh$ argument becomes small enough ($\approx 1$) the potential relief near the minimum flattens. Here the $\tanh$ argument is multiplied either by $1+\sigma$ or by $1-\sigma$. Thus, the minima for LS and HS in both layers are much more stable, and the minima with opposite spin states can be destroyed by large enough $\sigma$. Corresponding equation reads
\begin{equation}\label{scr}
  \frac{1-\sigma_c}{t_c^{(2)}} \approx 1 \Leftrightarrow \sigma_c \approx  1 - \frac{2 \delta}{\ln{g}}.
\end{equation}
It is also applicable if $(\delta_1-\delta_2)/(\delta_1+\delta_2) \ll 1$.

At $\sigma>\sigma_c$ barrier height at $t_c^{(2)}$ is defined with good accuracy by $f(+1,-1)-f(-1,-1)$ or $f(-1,+1)-f(-1,-1)$. It gives $h=2 \sigma - (\delta_2 - \delta_1)$ or $h=2 \sigma - (\delta_1 - \delta_2)$, correspondingly. At $\delta_1=\delta_2$ the minimal barrier height is $2 \sigma_c \approx 2 - 2 t_c^{(2)}$. Thus, at $t_c^{(2)} < 2/3$ the first order transition takes place. At $t_c^{(2)} > 2/3$ the character of the transition depends on $\sigma$.



As in the one-layer problem, by variation of $\delta$ in both layers the system can be switched from LS to HS state and vice versa.

However, we notice a new feature with respect to the one-layer problem. One can tune the splitting $\delta_1$ at one layer, due to the interaction with the substrate, keeping the splitting at the other one $\delta_2$ and the temperature $t$ fixed. This idea is illustrated in Fig.~\eqref{Fig2}, where the phase diagram of the system in $(\sigma,t)$ plane is mapped onto the corresponding phase diagram in $(\sigma,\delta_1)$ plane. Variation of $\delta_1$ around
\begin{equation}
\delta_{1,c}=t \ln g-\delta_2.
\end{equation}
allows one to switch from LS to HS and vice versa simultaneously in both layers, i.e. to induce a spin transition in one layer by its interaction with another layer.


\subsection{Multiple layers}

Let us now consider some general results for $n$ layers. The free energy in this case has the form:
\begin{equation}
  \begin{split}
  &f(m_1,..., m_n)=\frac{1}{2}(m_1^2+...+m_n^2+2\sigma m_1 m_2 + 2\sigma m_2 m_3 \\
  &+...+ 2 \sigma m_{n-1} m_n)\\
  &-t\bigl[\ln(ge^{(m_1 + \sigma m_2 -\delta_1)/t}+e^{(\delta_1-m_1- \sigma m_2)/t}) \\
  &+\ln(ge^{(m_2 + \sigma (m_1+m_3) -\delta_2)/t}+e^{(\delta_2-m_2- \sigma (m_1+m_3))/t}) +... \\
  &+ \ln(ge^{(m_{n-1} + \sigma (m_{n-2}+m_{n}) -\delta_{n-1})/t} \\ &+e^{(\delta_{n-1}-m_{n-1} - \sigma (m_{n-2}+m_{n})/t}) \\
  &+ \ln(ge^{(m_n + \sigma m_{n-1} -\delta_n)/t}+e^{(\delta_n-m_{n} - \sigma m_{n-1})/t})\bigr].
  \end{split}
\end{equation}
Similar to previous section calculations give minima at $(+1,...,+1)$ and $(-1,...,-1)$ with free energies
\begin{eqnarray}\label{mfen2}
  f(-1,...,-1) &\approx& - n/2 - (n-1) \sigma  - n \overline{\delta},  \\ \label{mfen3}
  f(+1,...,+1) &\approx&  - n/2 - (n-1) \sigma - n t \ln{g} + n \overline{\delta},
\end{eqnarray}
where $\overline{\delta}=(\delta_1+...\delta_n)/n$ is mean value of $\delta_\alpha$. From these equation we get
\begin{equation}\label{tcn}
  t^{(n)}_c=\frac{2 \overline{\delta}}{\ln{g}}.
\end{equation}

Conditions of other possible minima stability (they have the form $(\pm 1,...,\pm 1 )$) at $t^{(n)}_c$ are similar to the discussed above. They depend on the value of arguments in $\tanh$ functions:
\begin{eqnarray}
  && \tanh{\frac{m_1 + \sigma m_2}{t^{(n)}_c}},  \\
  && \tanh{\frac{m_{i} + \sigma (m_{i-1}+m_{i+1})}{t^{(n)}_c}}, \\
  && \tanh{\frac{m_{n} + \sigma m_{n-1}}{t^{(n)}_c}}.
\end{eqnarray}
Let's start from $(-1,...,-1)$ state. If we flip some layer in the middle we will have condition of this texture stability in form $(1-2 \sigma)/t^{(n)}_c > 1$. However textures with $(-1,...,-1,+1,...,+1)$ --- ``domain walls''--- are much more stable, the corresponding condition reads $1/t^{(n)}_c > 1$. Thus, we can estimate the barrier height at $t^{(n)}_c$ as $f(-1,...,-1,+1,...,+1)-f(-1,-1,...,-1)$. For the $(-1,...,-1,+1,...,+1)$ state the potential relief for layers at the ``domain wall'' are almost flat at $1/t^{(n)}_c > 1$ and the average spin of the corresponding layers is zero. So, we obtain for equal $\delta_i$ case:
\begin{equation}\label{mfen4}
  \begin{split}
  f(-1,...,-1,+1,...,+1) &\approx -\frac{n-2}{2} - (n-2)\delta  \\ & -(n-3)\sigma - \frac{4 \delta}{\ln{g}}\left( \ln{2\sqrt{g}}\right),
  \end{split}
\end{equation}
Thus, the barrier height reads
\begin{equation}\label{mbar1}
  h \approx 1 + 2\sigma -2 t^{(n)}_c \ln{2}.
\end{equation}
This quantity can be used for estimating whether we have the first order transition or smooth crossover.

As in the previous Subsec. \ref{bilayer} we notice a possibility of switching the spin state of the whole system by variation of $\delta_1$. However, it is seen from Eq.~\eqref{tcn} that the impact of $\delta_1$ has an additional factor $2/n$ in comparison with~\eqref{dt12}, which makes it rather weak for $n \gg 1$. Thus, the interaction of the first layer with the substrate, which is important in the double-layer problem, is almost negligible. So, similar to the one-layer problem, the spin state of the whole sample can be switched by the temperature or by the $\delta$ variation in all the layers, by means of e.g. external electric field. An important issue, which should be taken into account is the existence of the domain walls, which are metastable at low temperatures.

\section{Summary and conclusion}

To conclude, we theoretically address the problem of spin transitions in the systems consisting of molecular layers. In the framework of the mean-field approach we obtain the already established physics of this systems, which includes the first-order-like thermal phase transitions with hysteresis and smooth crossovers from the low spin state of the system to the high spin state. We further consider the possibility of isothermal switching by means of an electric field, which is provided by, e.g., ferroelectric substrate. For the bulk problem (single layer) we determine the conditions under which the hysteresis due to the variation of the energy splitting between the LS and HS states can appear. Experimental observation of such hysteresis would be a crucial step toward technological implementation of SCO films in nanoelectronics.

In the case of layered structures we find that two qualitatively different situations should be distinguished: the system consisting of few layers ($n \sim 1$) and multilayer systems with $n\gg1$. In both cases it should be possible to observe a staircase in the total magnetization as a function of the electric field in a certain range of parameters. We call such phenomenon \textit{multistability}. For $n \sim 1$, provided the interlayer coupling is sufficiently large, all the layers can be switched simultaneously by switching only the first layer by, e.g., microscopic interaction with the surface of the substrate. We believe this effect to be relevant to the experimental findings \cite{Zhang1, Zhang2}.  In contrast, multilayer systems with $n\gg1$ behave analogously to the bulk: the boundary plays no role and the total magnetization of the film can be controlled either by temperature or by the macroscopic electric field produced by the substrate. The latter must be sufficiently strong for the coupling \eqref{Coupling} could come into play. One should also take care of highly stable intermediate states --- ``domain walls'' --- which should be avoided in the switching process. Detailed investigation of this phenomenon is beyond the scope of this paper, and will be given elsewhere.

\begin{acknowledgments}

S. V. thanks A. Cano for helpful discussions and acknowledges the financial supports by CNRS and by Russian Science
Foundation (Grant No. 18-72-00013). Contribution to the work by O. I. Utesov was funded by RFBR according to the research project 18-02-00706.

\end{acknowledgments}

\end{document}